\documentclass[aps,prl,twocolumn,showpacs,groupedaddress,amsmath,amssymb]{revtex4}

\usepackage{ulem}
\usepackage{graphicx}
\usepackage{color}
\usepackage{comment}
\usepackage{bm}
\usepackage{latexsym}
\usepackage[latin9]{inputenc}
\usepackage{amsmath}
\usepackage{esint}
\usepackage{braket}

\usepackage[english]{babel}

\begin{document}

\title{Floquet Perfect Absorbers based on Periodically Driven Targets}
\author{Dashiell Halpern$^1$, Suwun Suwunnarat$^1$, Huanan Li$^{1}$\footnote{hli01@wesleyan.edu}, 
Boris Shapiro$^{2}$\footnote{boris@physics.technion.ac.il}, Tsampikos Kottos$^{1}$\footnote{tkottos@wesleyan.edu}} 
\affiliation{$^1$Wave Transport in Complex Systems Lab, Department of Physics, Wesleyan University, Middletown, CT-06459, USA}
\affiliation{$^2$Technion - Israel Institute of Technology, Technion City, Haifa 32000, Israel}
\date{\today}

\begin{abstract}
We introduce the concept of multichannel Floquet Perfect Absorbers (FPAs) which are periodically modulated lossy interferometric 
traps that completely absorb incident monochromatic waves. The proposed FPA protocols utilize a Floquet engineering approach  
which inflicts a variety of emerging phenomena and features: {\it reconfigurability} of perfect absorption (PA) for a broad range of 
frequencies of the incident wave; PA for infinitesimal local losses, and PA via critical coupling with high-Q modes by inducing back-
reflection {\it dynamical} mirrors. 
\end{abstract}

\maketitle

{\it Introduction --}The quest of new methods and technologies that can lead to a perfect absorption (PA) of an incident waveform is 
an interdisciplinary research theme in classical wave physics.  It spans a range of frequencies from optics \cite{WLP12,CGCS10,DR12,
L10,WCGNSC11,CS11,ZMZ12,STLLC14,PF14,KS14,VBPA15,BBFSN15,SCCWBDPCMS16} and microwaves \cite{S50,LSMSP08,
PL10,PPVZRKL13}, to radio-frequencies \cite{DK38,S52}, acoustics \cite{MMYYWS12,MYXYS14,SBHL14,RTRMTP16} and electronics 
\cite{CGMM13,PCZW13,SLLREK12}. A successful outcome can revolutionize a variety of wave physics applications including energy
conversion \cite{LGJSY05,T07} and photovoltaics \cite{AP10,LCG06,LH08}, imaging techniques \cite{FCDPRTTW00,BPTB02,MRDNF02,
CPCLD04} and medical therapies \cite{HRY15}, stealth technologies \cite{S52,FM88,VJ96} and soundproofing \cite{M12,MYXYS14}. 

A desirable feature for many of the above applications is an ``on-the-fly" reconfigurability of the structure i.e. the possibility to absorb 
on demand an incoming monochromatic wave at a specific frequency, without altering the fabrication characteristics of the structure
itself. Another requirement, either due to cost or design considerations, is to incorporate minimal losses inside the structures while at 
the same time achieve a perfect absorption. This second requirement has been recently addressed, in the frame of multi-channel 
systems, by the so-called {\it Coherent Perfect Absorber} (CPA) scheme. This is an interferometric protocol employing two counter-propagating 
waves which destructively interfere outside a weakly lossy cavity (the target)-- in analogous manner to time-reversed laser - to achieve 
coherent perfect absorption \cite{CGCS10,WCGNSC11}. The original concept, involving only two channels coupled to a simple cavity 
has been realized in various frameworks \cite{WCGNSC11,SLLREK12,STLLC14,ZFYZGSF16} and further extended to include multi-
channel complex cavities \cite{LSFSK17,FSK17,LSK17}. CPA is a ``generalization" of an older scheme, applicable only for single-channel 
cavities, which employs a back-reflection mirror and {\it critical coupling} to the resonances of the lossy cavity. Unfortunately, two (or multi)
-sided coherent wave injection in some applications might be challenging to implement, while back-reflection mirrors are often either lossy 
(e.g. metallic mirrors) or require additional fabrication effort (e.g. distributed Bragg reflectors). To make things worst, none of the above 
proposals addresses the ``on-the-fly" reconfigurability issue. Obviously, a more flexible scheme that can adopt one of the other approach, 
depending on the application at hand, is extremely desirable. 

\begin{figure}
\includegraphics[width=1\linewidth, angle=0]{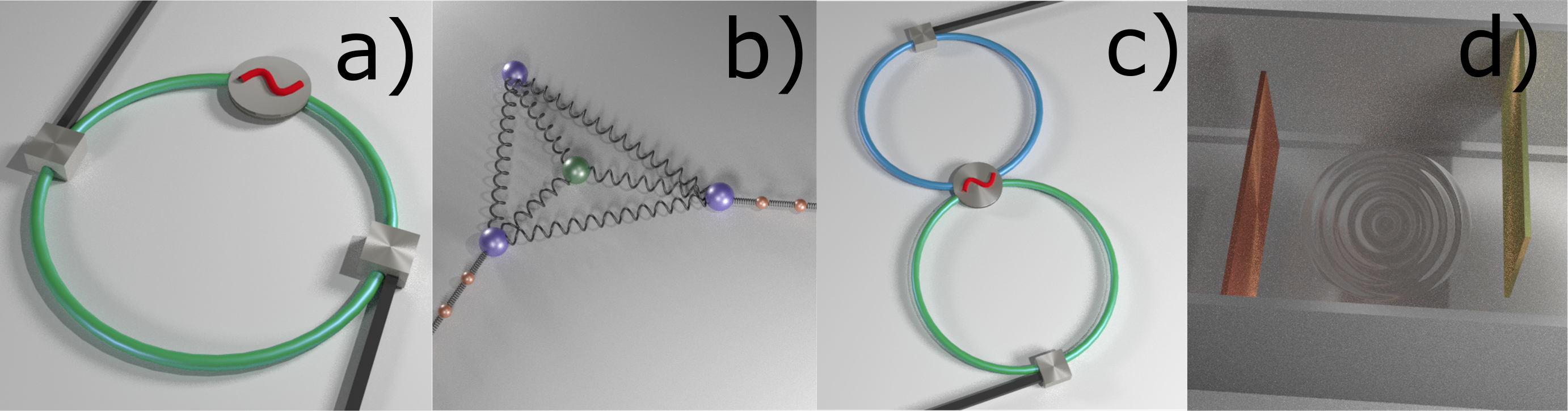}
\caption{(Color Online) Schematics of various physical set-ups that can be used for FPA: (a) A single-mode fiber coupled to a time-periodic 
phase modulator; (b) A network of four oscillators (resonators) with modulated coupling constants; (c) Two single-mode fibers with a time-
modulated coupler; and (d) an acoustic cavity with two cantilevers where the in-between air domain is periodically modulated in time. Weakly 
lossy elements are indicated with green.  
}
\label{fig1}
\end{figure}

In this paper we introduce PA in a completely different framework associated with Floquet time-modulated cavities. Using a 
Floquet scattering formalism we show that the proposed {\it Floquet Perfect Absorber} (FPA) protocols are ``on-the-fly" reconfigurable 
and can PA any specified frequency of the incoming waves by altering the external driving characteristics --  a feature that is absent 
from any static PA scheme whose properties are fixed during the fabrication process. Importantly, we show that the implementation 
of Floquet engineering methods provide powerful means for the implementation of PA protocols with unconventional characteristics. 
For example, we demonstrate perfect absorption via coherent multi-channel illumination for {\it infinitesimal local} losses -- a 
phenomenon which we refer to as {\it Floquet-enhanced perfect absorption} -- occurring at appropriate values of the frequency/
amplitude of the periodic driving scheme. Finally, we show that our scheme can be flexible and interpolate between coherent FPA
and PA via critical coupling associated with single-channel illumination. In contrast to conventional wisdom, we demonstrate that the
latter scenario can occur even in the absence of physical back-reflecting mirrors. The latter is induced dynamically via Floquet engineering.

{\it Floquet Scattering Formalism --} We consider a cavity, or a network of $N_s$ coupled single-mode cavities. Hereafter we refer 
to the cavities as ``sites" with the corresponding resonant mode attached to each site. The mode frequencies, and generally the couplings, 
are periodically modulated by an electric field which is periodic in time with period $T=2\pi/\omega$. For some
physical set-ups see Fig. \ref{fig1}. In the context of coupled-mode theory, the time-periodic Floquet resonator network can be described 
by an effective time-periodic Hamiltonian $H\left(t\right)=H\left(t+2\pi/\omega\right)$ which takes the form
\begin{align}
H\left(t\right) & =H_{0}\left(t\right)-\imath\Gamma,\quad\Gamma=\sum_{\mu}\gamma_{\mu}\left|e_{\mu}\right\rangle \left\langle 
e_{\mu}\right|.
\label{H_F}
\end{align}
Above $H_{0}\left(t\right)$ is a $N_{s}\times N_{s}$ Hermitian matrix, $\gamma_{\mu}$ quantifies the loss in the $\mu$-th cavity and 
$\left\{ \left|e_{\mu}\right\rangle \right\}$ is the basis of the mode space where $H_{0}\left(t\right)$ is represented. We turn the system 
of Eq. (\ref{H_F}) to a scattering set-up by attaching to it two static semi-infinite leads $\alpha = L, R$, each of which is supporting
plane waves with a dispersion relation $E(k)$. We shall assume, for demonstration purposes, that the leads feature an one-dimensional 
tight-binding dispersion $E=-2\cos k$ (in units of coupling) and that the loss strength of the lossy resonator(s) are $\gamma_{\mu}=
\gamma$. 

When an incident wave with frequency $E_0=E(k_0)\in [-\omega/2,\omega/2]$ is engaged with the periodically-modulated target, it is 
scattered to an infinite number of outgoing channels (including evanescent ones) supporting frequencies $E_{n}=E_0+n\omega=-2
\cos k_{n}$ ($n$ is an integer). The evanescent channels in the leads with $E_{n}\notin\left(-2,2\right),\mathrm{Im}k_{n}>0$ do not carry 
flux. We, therefore, consider the scattering matrix $\mathcal{S}$ which connects only the $N_p$ incoming with the outgoing 
propagating channels $E_{n}\in\left(-2,2\right),k_{n}\in\left(0,\pi\right)$ at each of the $\alpha=L,R$ leads. Following Ref. \cite{Li2018} 
we write the flux-normalized scattering matrix $S$ as 
\begin{align}
\mathcal{S} & =-I_{2N_{p}}+\imath WG_{s}W^{T},G_{s}\equiv\frac{1}{E_0-H_{Q}+W_{c}^{T}KW_{c}}
\label{scattering_matrix}
\end{align}
where $I_{2N_{p}}$ is the $2N_{p}\times2N_{p}$ identity matrix. The quasi-energy operator $H_{Q}$ is defined in the Floquet-Hilbert
space and has elements $\left(H_{Q}\right)_{ns,n's'}=H_{ss'}^{\left(n-n'\right)}-\delta_{nn'}n\omega$ where the Fourier components are 
$H_{ss'}^{\left(n\right)}\equiv{1\over T}\int_{0}^{T}dtH_{ss'}\left(t\right)\exp\left(\imath n\omega t\right)$ and $n,n'$ are integers while 
$s$ labels the sites (resonators) of the system. $W$ is the coupling matrix that describes the coupling between the 
propagating-channels (at the leads) and the system. Its matrix elements are $\left(W\right)_{n_{P}\alpha,n's}=\sqrt{v_{n_{P}}}c_{\alpha}
\delta_{n_{P}n'}\delta_{\alpha\leftrightarrow s}$ where $v_{n_{P}}\equiv\left.\partial E/\partial k\right|_{k_{n_{P}}}=2\sin k_{n_{P}}$
and the sub-index $n_{P}$ labels only the propagating channels. The matrix $\left(W_{c}\right)_{n\alpha,n's}=c_{\alpha}\delta_{nn'}
\delta_{\alpha\leftrightarrow s}$ and $c_{\alpha}$ describes the bare coupling between the lead $\alpha$ and the Floquet system
where we define $\delta_{\alpha\leftrightarrow s}=1$ when lead $\alpha$ is coupled with the site s directly or $\delta_{\alpha\leftrightarrow 
s}=0$ otherwise. Finally the matrix $K$ takes the form $\left(K\right)_{n\alpha,n'\alpha'}=\delta_{nn'}\delta_{\alpha\alpha'}
\exp\left(\imath k_{n}\right)$.

{\it Necessary Conditions for Floquet Perfect Absorption --} The scattering matrix $\mathcal{S}(E_0,\gamma,\omega)$, 
Eq.~(\ref{scattering_matrix}), relates the incoming wave (in the propagating channel representation) $\left|{\cal I}\rangle\right.$ 
to an outgoing wave $\left|{\cal O}\rangle\right.$ emerging after the scattering with the periodically time-modulated 
target. In other words we have that $ \mathcal{S}(E_0,\gamma,\omega)\left|{\cal I}\rangle\right.=\left|{\cal O}\rangle\right.$. The 
condition for perfect absorption follows by requiring that the outgoing wave is the null vector, i.e. $\left|{\cal O}\rangle\right.=0$. 
The latter is satisfied for a set of {\it real valued} scattering parameters $(E_{\rm FPA},\gamma_{\rm FPA},\omega_{\rm FPA})$ 
for which $\det \left[\mathcal{S}\left(E_{\rm FPA},\gamma_{\rm FPA},\omega_{\rm FPA}\right)\right]=0$. While the reality of the 
driving frequency $\omega$ and the loss-strength parameter $\gamma$ are dictated by the formulation of the problem itself, the 
requirement for real incident frequencies $E_0=E_{\rm FPA}$ is based on physical considerations; namely the fact that the 
incoming wave has to be a propagating wave. Using Eq.~(\ref{scattering_matrix}) we are able to recast the above condition for 
FPA to the following form
\begin{align}
\det\left(E_{FPA}-H_{Q}+W_{C}^{T}KW_{c}-\imath W^{T}W\right) & =0
\label{secular}
\end{align}
which resembles a generalized eigenvalue problem associated with an effective non-Hermitian Hamiltonian $H_{\rm eff}= H_{Q}-W_{C}^{T}
KW_{c}+\imath W^{T}W$. 


In the small coupling limit $c_{\alpha}\rightarrow0$, a first-order perturbation approach allows us to evaluate theoretically $\left(E_{FPA},
\gamma_{FPA}\right)$ from Eq.~(\ref{secular}) \cite{note0}. We have,
\begin{align}
E_{FPA} & \approx E^{\left(0\right)}-\sum_{n_{E},\alpha}c_{\alpha}^{2}\left|\psi_{n_{E}s_{\alpha}}^{\left(0\right)}\right|^{2}e^{\imath 
k_{n_{E}}^{\left(0\right)}}-\sum_{n_{P},\alpha}c_{\alpha}^{2}\left|\psi_{n_{P}s_{\alpha}}^{\left(0\right)}\right|^{2}\cos k_{n_{P}}^{\left(0\right)}
\nonumber \\
\gamma_{FPA} & \approx\sum_{n_{P}\alpha}c_{\alpha}^{2}\left|\psi_{n_{P},s_{\alpha}}^{\left(0\right)}\right|^{2}\sin k_{n_{P}}^{\left(0\right)}
/\sum_{n,\mu}\left|\psi_{n,\mu}^{\left(0\right)}\right|^{2}
\label{CPA_pair}
\end{align}
where $\psi_{ns}^{\left(0\right)}=\left\langle e_{ns}\right.\left|\psi^{\left(0\right)}\right\rangle $, $\left|e_{ns}\right\rangle $ is the unit vector 
in the Floquet-Hilbert space with the entry being $\left(\left|e_{n\mu}\right\rangle \right)_{n's}=\delta_{nn'}\delta_{\mu s}$ and $\left\{ E^{
\left(0\right)},\left|\psi^{\left(0\right)}\right\rangle \right\} $ is an eigenpair of the Hermitian matrix $H_{Q}\left(\gamma=0\right)=H_{Q}^{\dagger}
\left(\gamma=0\right)$, i.e., $H_{Q}\left(\gamma=0\right)\left|\psi^{\left(0\right)}\right\rangle =E^{\left(0\right)}\left|\psi^{\left(0\right)}\right
\rangle$ and $\left\langle \psi^{\left(0\right)}\right.\left|\psi^{\left(0\right)}\right\rangle =1$. The index $\mu$ indicates the lossy resonators, 
$n_{E}$ indicates the evanescent channels and $n_{P}$ the propagating channels. Finally $s_{\alpha}$ labels the resonators which are 
coupled with the lead $\alpha$ directly, and $k_{n}^{\left(0\right)}$ is obtained from the dispersion $E_{n}^{\left(0\right)}=E^{\left(0\right)}
+n\omega=-2\cos k_{n}^{\left(0\right)}$.

{\it Coherent FPA scheme--}Equation (\ref{secular}), and its perturbative variant Eq. (\ref{CPA_pair}), are necessary conditions 
for PA. In case of multi-channel targets, however, one needs to impose an additional constraint; the incident waveform $\left|{\cal I}
\rangle\right.$ must be a linear combination of channel modes with amplitudes given by the components of the eigenvector $\left|
{\cal I}_{\rm FPA}\rangle\right.$ of the scattering matrix Eq. (\ref{scattering_matrix}) associated with a zero eigenvalue $s_{\rm FPA}
(E_{\rm FPA},\gamma_{\rm FPA},\omega_{\rm FPA})=0$. Such coherent incident waveform induces interference that trap the wave 
inside the structure, thus leading to a complete absorption. We refer to this scenario as Coherent FPA.

As a useful illustration, we solve the FPA problem explicitly in the case of one driven lossy resonator, i.e., $H_{0}\left(t\right)=h\left(t\right),
h\left(t\right)\in R$, see Fig. \ref{fig1}a. Generally the eigenvectors $\left|\psi_{n}^{\left(0\right)}\right\rangle$ of the quasi-energy operator 
$H_{Q}\left(\gamma=0\right)$ are related to the Floquet mode associated with the Hamiltonian $H_{0}\left(t\right)$. In the case of one 
driven resonator, we can write the eigenvalues $E_{n}^{\left(0\right)}$ and the corresponding eigenvectors $\left|\psi_{n}^{\left(0\right)}
\right\rangle$ of $H_{Q}\left(\gamma=0\right)$ explicitly using the driving $h\left(t\right)$. Specifically, we have 
\begin{align}
E_{n}^{\left(0\right)} & =E^{\left(0\right)}+n\omega,\quad E^{\left(0\right)}=\frac{\omega}{2\pi}\int_{0}^{2\pi/\omega}dth\left(t\right)\nonumber \\
\left\langle n'\right.\left|\psi_{n}^{\left(0\right)}\right\rangle  & =\frac{\omega}{2\pi}\int_{0}^{2\pi/\omega}dte^{\imath\left(n'+n\right)\omega t}u
\left(t\right)
\label{eq: eigensolution}
\end{align}
where $u\left(t\right)=\exp\left(-\imath\int_{0}^{t}dt'h\left(t'\right)+\imath E^{\left(0\right)}t\right)u\left(0\right)$ is the Floquet state of the time-
periodic Hamiltonian $h\left(t\right)$ with the initial condition being $u\left(0\right)$.

In order to make further progress, we now consider a specific example where $h\left(t\right)=\beta\cos\omega t$. Using Eq.~\eqref{eq: eigensolution}, 
we obtain $E_{n}^{\left(0\right)}=n\omega$ and $\left\langle n'\right.\left|\psi_{n}^{\left(0\right)}\right\rangle =J_{n'+n}\left(\beta/\omega\right)u
\left(0\right)$, where $J_{n'+n}$ denotes the Bessel function of first kind of the integer order $n'+n$. Under the assumption that there exists 
only one propagating channel $E$ in each lead, we get from Eq.~(\ref{CPA_pair})
\begin{align}
E_{FPA} & \approx0,\;\gamma_{FPA}\approx\left(c_{L}^{2}+c_{R}^{2}\right)J_{0}^{2}\left(\beta/\omega\right)\label{eq: example}
\end{align}
From Eq.~\eqref{eq: example}, we see that due to the factor $J_0^2$, the lossy strength $\gamma_{FPA}$ required for the realization 
of FPA can be dramatically reduced. We refer to this phenomenon as Floquet-enhanced PA. 

\begin{figure}
\includegraphics[width=1\linewidth, angle=0]{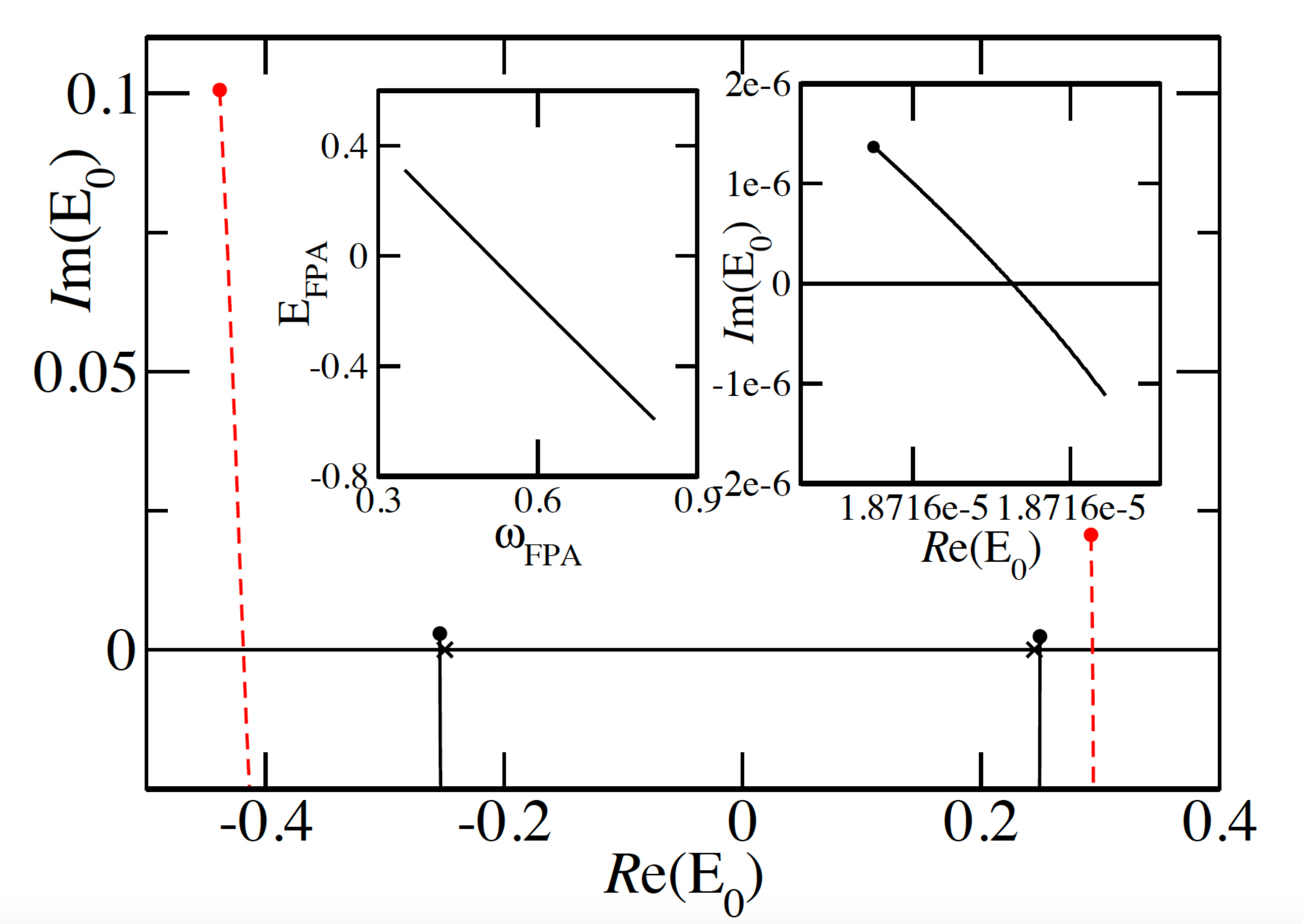}
\caption{(Color Online) Parametric evolution of the zeros of the $\mathcal{S}$-matrix as the loss $\gamma$ at one resonator of a 
network of four coupled resonators, increases (see supplement for the corresponding Hamiltonian and Fig. \ref{fig1}b for a mechanical
analogue). The driving frequency is $\omega=1$ and the amplitude is $\beta=0.9$. The position of the zeroes for $\gamma=0$ are 
indicated with filled circles. The two set of trajectories correspond to strong $c\approx -1$ (red-dashed lines) and weak $c\approx -0.2$ 
(black- solid lines) coupling of the resonators with the leads. The crosses indicate the predictions of perturbation theory Eq. (\ref{CPA_pair}). 
Right inset: The FPA can be reconfigured to occur for an extremely small value of $\gamma_{\rm FPA}\approx 2.76\times 10^{-6}$ 
when the driving frequency is $\omega=1$, and the driving amplitude is $\beta=0.01$ ($c=-0.5$). Left inset: The Floquet 
network (with fixed topology and loss $\gamma=0.0322$) can be reconfigured by changing the driving frequency $\omega$ and amplitude 
$\beta$ in order to perfectly absorb an incident wave with a dense set of frequencies $E_0$. We show the numerically evaluated 
$E_{\rm FPA}$ versus $\omega_{\rm FPA}$.  
}
\label{fig2}
\end{figure}

More complicated systems can be also used for the implementation of the coherent FPA scheme. Take for example a network of coupled 
resonators or oscillators like the one shown schematically in Fig. \ref{fig1}b. This system consists of four coupled resonant modes 
with the central one having losses $\gamma$. For 
the numerical demonstration we have assumed that the coupling strength between the resonators can be modulated in a sinusoidal 
manner (the Hamiltonian $H(t)$ that describes the isolated driven system is given in the supplement). In Fig. \ref{fig2} we report the 
parametric evolution of the complex zeroes of the scattering matrix found from Eq. (\ref{secular}) for two different sets of coupling constants. 
For $\gamma=0$, the complex zeroes $E$ are lying on the upper complex frequency semi-plane as a consequence of causality. 
When, however, $\gamma\neq 0$ these eigenvalues can, in principle, situated in both positive and negative half-planes of the complex 
frequency plane. The ones that have crossed the real axis at frequency $E_{\rm FPA}$ are relevant to our study, since for the 
corresponding loss-strength and/or driving frequency $\gamma_{\rm FPA}, \omega_{\rm FPA}$ (and driving amplitude $\beta$) 
an incident traveling waveform can be perfectly absorbed. In Fig. \ref{fig2} we also mark with crosses (in the real $E$-axis) the 
theoretical predictions Eq. (\ref{CPA_pair}) associated to the case of weak coupling between the system and the leads. Furthermore,
at the right inset of Fig. \ref{fig2} we show 
the parametric evolution of the zeros in case that a Floquet-enhanced PA is engineered via a choice of appropriate
frequency and amplitude of the modulated target. Specifically, for modulation frequency $\omega=1$ and amplitude $\beta=0.01$ 
we have observed a crossing of the zeros with the real axis which occurs for loss-strength as low as $\gamma= 2.7601\times 
10^{-6}$ (in coupling units).

An important element of our Floquet perfect absorption protocol is the possibility to induce PA at different incident frequencies $E_0
=E_{\rm FPA}$ without changing the fabrication characteristics of the cavity. Numerical evaluation of the secular Eq. (\ref{secular}) 
for an example case of a network of four coupled resonators (see supplement), indicates that these Floquet cavities can be reconfigured 
to act as PAs for a dense set of incident frequencies $E_0$  by simply changing the driving frequency $\omega=\omega_{\rm FPA}$ 
(and driving amplitude $\beta$) of the coupling modulation, see left inset of Fig. \ref{fig2}.

Let us finally comment on an alternative formulation that allows us to evaluate the coherent FPA values $(E_{\rm FPA},\gamma_{\rm FPA})$ 
together with the corresponding incident waveform $\left|I_{FPA}\right\rangle$. This approach involves the notion of the absorption 
matrix $A(E,\gamma,\omega)\equiv I_{2N_{p}}-S^{\dagger}S=A^{\dagger}$. The eigenvalues $\alpha(E,\gamma,\omega)$ of the 
absorption operator indicate the amount of absorption that a coherent 
incident waveform, with channel amplitudes dictated by the components of the associated eigenvector, will experience once it encounter the 
modulated target. Obviously the eigenvalues of the absorption operator are $0 \leq \alpha\leq 1$; when $\alpha(E,\gamma)=0$ the incident 
waveform is not absorbed, while $\alpha(E=E_{\rm FPA},\gamma=\gamma_{\rm FPA})=1$ indicates complete absorption. Using Eq.~
(\ref{scattering_matrix}), we can re-write the absorption matrix $A$ in a simpler form
\begin{align}
A & =2\gamma\sum_{n}\sum_{\mu}\left|u_{n\mu}\right\rangle \left\langle u_{n\mu}\right|,\left|u_{n\mu}\right\rangle \equiv WG_{s}^{\dagger}
\left|e_{n\mu}\right\rangle .
\label{absoption}
\end{align}
In the weak-coupling limit and $E_0=E_{FPA}, \gamma=\gamma_{FPA}$, we have $G_{s}\approx\left\langle \psi^{\left(0\right)}\right|G_{s}
\left|\psi^{\left(0\right)}\right\rangle \left|\psi^{\left(0\right)}\right\rangle \left\langle \psi^{\left(0\right)}\right|$ and thus $\left|I_{FPA}\right
\rangle \propto$$W\left|\psi^{\left(0\right)}\right\rangle$. Therefore, the study of coherent FPA in the weak-coupling limit boils down to the 
eigenvalue problem of the operator $H_{Q}\left(\gamma=0\right)$.

{\it FPA based on Dynamical Mirrors and Critical Coupling  --} The implementation of PA protocols that do not need a coherent multi-sided 
illumination or a back-reflection mirror, is highly attractive for many applications and could open up many engineering possibilities. One way 
to achieve this goal is by utilizing the presence of accidental degeneracies of critically coupled modes with opposite symmetries \cite{PLF14}
 -- a quite demanding scheme in terms of organizing appropriately resonant modes and their quality factors. Below we propose an altogether 
different approach which utilizes critical coupling to resonances. In this case, the physical back-reflection mirrors are absent and instead, we 
utilize appropriate Floquet driving schemes that generate {\it dynamical} mirrors. 

The basic idea can be demonstrated using the simple system of Fig. \ref{fig1}c, described by the effective Hamiltonian Eq. (\ref{H_F}) with 
$H_{0}\left(t\right)=\begin{pmatrix}\varepsilon_{L} & -e^{\imath\omega t}\\
-e^{-\imath\omega t} & \varepsilon_{R}
\end{pmatrix}$ and $\Gamma=\begin{pmatrix}0 & 0\\
0 & \gamma
\end{pmatrix}$ due to a ``right'' lossy resonator. To this end, we consider that an incident wave with energy $E_0$, impinges the driven target from  
the left lead. The driving will couple the propagating channel $E_{0}$ only with the $E_{1}=E_0+\omega$ channel in the right lead \cite{LKS18}. To 
realize an FPA in this framework, we consider $E$ and $\omega$ values such that $E_{1}>2$ (band edges) corresponding to an evanescent 
channel carrying zero flux. Essentially in this scenario, the Floquet scheme generates an impenetrable wall (dynamical mirror) at the right lead 
which enforces total reflection of the impinging wave. Consequently, the reflected wave exits the scattering domain at the same frequency $E_0$, 
as the incident wave. Using Eq.~(\ref{scattering_matrix}), we obtain the reflection amplitude (in case of perfect coupling $c=-1$)
\begin{align}
r_{0} & =-\frac{1-\left(\varepsilon_{L}+e^{\imath k_{0}}\right)\left(\varepsilon_{R}-\imath\gamma+e^{-\imath k_{1}}\right)}{1-\left(\varepsilon_{L}
+e^{-\imath k_{0}}\right)\left(\varepsilon_{R}-\imath\gamma+e^{-\imath k_{1}}\right)}.
\label{eq: r0}
\end{align}
Using Eq. (\ref{eq: r0}) together with the FPA condition $r_{0}=0$ we can evaluate the FPA points $(E_{\rm FPA},\gamma_{\rm FPA},
\omega_{\rm FPA})$. 

\begin{figure}
\includegraphics[width=1\linewidth, angle=0]{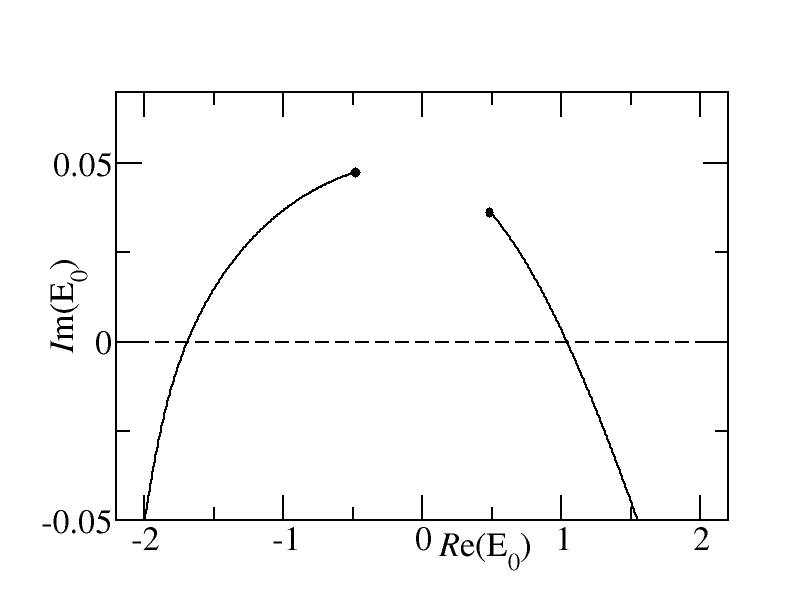}
\caption{(Color Online) Parametric evolution (versus increasing driving amplitude $\beta$) of the zeros of the $\mathcal{S}$ matrix 
associated with a network of four fully connected resonators, two of which are coupled to leads (see Supplement). The other two 
resonators have losses characterized by a loss-strength $\gamma=0.4749$ (fixed in these simulations). The driving frequency of the 
coupling is $\omega=5$ and the amplitude is increased from $\beta=0.15$ (indicated with filled circle) to $\beta=0.3$. The FPA occurs 
for $\beta\approx 2.3$. This Floquet driving induces a dynamical mirror which can be used for PA via critical coupling.
}
\label{fig3}
\end{figure}

For example, for an incident wave at frequency $E_0=E_{\rm FPA}=0$ (middle of the band) we find that an FPA occurs at loss strength 
$\gamma_{\rm FPA}\approx 1/\varepsilon_{L}^2$ and driving frequency $\omega_{\rm FPA}\approx 2\left(\varepsilon_{R}-\sqrt{
\gamma_{\rm FPA}}\right)$ \cite{note1}. These FPA points have a simple physical interpretation: In the limit of small losses $\gamma_{
\rm FPA}$, the on-site potential $\epsilon_L\sim (1/\sqrt{\gamma})$ at the left site has to take large values. At the same time the Floquet 
driving creates a dynamical wall that forbids the incident wave to escape from the right 
lead. In other words, the scattering system turns to a high-Q cavity. When the incoming wave is at resonant with the modes of the cavity 
(i.e. perfect impedance matching) then it can be trapped for large times and eventually absorbed completely -- even if the absorption 
strength $\gamma$ is infinitesimally small. The above scenario is nothing else than the so-called impedance matching condition, which, 
once expressed in terms of losses, indeed states that radiative and material losses must be equal \cite{H84}.

The Floquet-induced critical coupling scenario of PA can be realized for more complicated cavities like a network of four fully connected 
resonators (see Fig. \ref{fig1}b). The model Hamiltonian that has been used in these simulations is shown in the supplement and assumes
that two of the resonators have local losses $\gamma$. In Fig. \ref{fig3} we show the parametric evolution of two of the complex zeroes 
of the scattering matrix $\mathcal{S}$ as the driving amplitude $\beta$ increases. We find that there is an FPA value at $\beta_{\rm FPA}$ 
for a specific driving frequency $\omega_{\rm FPA}$ and loss strength $\gamma_{\rm FPA}$ for which the zeroes cross the real axis. We, 
therefore, conclude that at this real frequency an incident wave exists which is completely absorbed by the network.

{\it Conclusions -} We have introduced a class of perfect absorbers that rely on Floquet engineering schemes. These Floquet Perfect 
Absorbers (FPA's) are easily reconfigurable, and can host a variety of new phenomena, including PA in the presence of infinitesimal 
local losses, and unidirectional reconfigurable PA's that operate without the use of physical mirrors. The latter are now substituted by 
dynamical (reconfigurable) mirrors realized using appropriate Floquet drivings schemes. Floquet PA's and other Floquet photonic systems 
is an emerging field which currently is in it's infancy. Among the promising future directions is the implementation of our FPA's proposal 
in existing Floquet platforms (see for example \cite{CLEK17}) and the possibility of extending the current study to the time-reversal 
scenario of FPA's i.e Floquet amplifiers and unidirectionally reconfigurable laser sources. This will be the theme of future investigations.


\begingroup
\clearpage
\onecolumngrid
\appendix

\begin{center}
\textbf{\large Supplemental Materials}
\end{center}

\setcounter{equation}{0}
\setcounter{figure}{0}
\setcounter{table}{0}
\setcounter{page}{1}
\makeatletter
\renewcommand{\theequation}{S\arabic{equation}}
\renewcommand{\thefigure}{S\arabic{figure}}
\renewcommand{\bibnumfmt}[1]{[S#1]}
\renewcommand{\citenumfont}[1]{S#1}

\section{Network system associated with Figure 2}

A network of coupled resonators has been used for the data shown in Figure 2. The network consists of four coupled resonators
with time-modulated couplings (for a mechanical analog of such network see Fig. 1b). The corresponding isolated system is described 
by the following effective time-dependent Hamiltonian $H(t)$:
\begin{equation}
H (t) = \left(\begin{array}{cccc}
  0 & - 1 + \beta \sin (\omega t) & - 1 + \beta \cos (\omega t) & - 1 + \beta
  \sin (\omega t)\\
  - 1 + \beta \sin (\omega t) & 0 & - 1 + \beta \cos (\omega t) & - 1 + \beta
  \sin (\omega t)\\
  - 1 + \beta \cos (\omega t) & - 1 + \beta \cos (\omega t) & - i \gamma & - 1
  + \beta \cos (\omega t)\\
  - 1 + \beta \sin (\omega t) & - 1 + \beta \sin (\omega t) & - 1 + \beta \cos
  (\omega t) & 0
\end{array}\right)
\label{H4}
\end{equation}
where we have assumed that there are two types of (out-of-phase) driving couplings $- 1 + \beta \sin (\omega t)$ and 
$- 1 + \beta \cos (\omega t)$. 

In order to study the FPA phenomena discussed in the main text, we have coupled this system with one-dimensional leads of
coupled resonators. The left lead is coupled directly with the first site (resonator) while the right lead is directly coupled to the
fourth site (resonator). Both (bare) couplings $c$ are assumed to be equal. The loss, with loss-strength $\gamma$, has been 
included in the third resonator.

The data shown in the main panel of Fig. 2 correspond to the following parameters: $\omega = 1$, $\beta = 0.9$ for  
two different couplings $c = 1$ (strong coupling) and $c=0.2$ (weak coupling). In this case the loss-strength $\gamma$
was increasing in order to obtain the parametric evolution of the zeros of the $\mathcal{S}$ matrix.

The data associated with the right inset of Fig. 2 correspond to driving parameters $(\beta=0.01; \omega=1)$ and coupling constant 
$c=0.5$. Finally the data associated with the left inset of Fig. 2 correspond to $\gamma_{\rm FPA} =0.0322; c = 0.5$ and driving amplitude 
$\beta$ ranging between $[0.1781, 0.472]$ with $\omega_{\rm FPA}$ and $E_{\rm FPA}$ varying as shown.

\

\section{Network system associated with Figure 3}

The results for Figure 3 are obtained when considering the time dependent Hamiltonian $H(t)$:

\begin{equation}
H (t) = \left(\begin{array}{cccc}
  0 & \beta \cos (2 \omega t) & \beta \cos (\omega t) & \beta \cos (3 \omega
  t)\\
  \beta \cos (2 \omega t) & - i \gamma & \beta \cos (\omega t) & \beta \cos
  (\omega t)\\
  \beta \cos (\omega t) & \beta \cos (\omega t) & - i \gamma & \beta \cos (2
  \omega t)\\
  \beta \cos (3 \omega t) & \beta \cos (\omega t) & \beta \cos (2 \omega t) &
  0
\end{array}\right)
\end{equation}
where $\gamma=0.4749, \omega = 5$ while the driving amplitude has been increased from $\beta =$0.15 to
0.3. The coupling to the leads is the same as the one used in the modeling for Fig. 2, with a coupling constant which is $c=0.25$.

\endgroup


\begin{thebibliography}{1}

\bibitem{WLP12}C. M. Watts, X. Liu, W. J. Padilla, Adv. Lett. {\bf 24}, OP98 (2012).

\bibitem{DR12}G. Dayal, S. A. Ramakrishna, Opt. Express {\bf 20}, 17503 (2012).

\bibitem{CGCS10}Y. D. Chong, L. Ge, H. Cao, A. D. Stone, Phys. Rev. Lett. {\bf 105}, 053901 (2010).

\bibitem{L10}S. Longhi, Physics {\bf 3}, 61 (2010).

\bibitem{WCGNSC11} W. Wan, Y. Chong, L. Ge, H. Noh, A. D. Stone, H. Cao, Science {\bf 331}, 889 (2011).

\bibitem{CS11}Y. D. Chong, A. D. Stone, Phys. Rev. Lett. {\bf 107}, 163901 (2011).

\bibitem{ZMZ12}J. F. Zhang, K. F. Macdonald and N. I. Zheludev, Light: Science and Appl.  {\bf 1}, 18 (2012).

\bibitem{PF14} J. R. Piper, S. Fan, ACS Photonics {\bf 1}, 347 (2014).

\bibitem{KS14} O. Kotlicki, J. Scheuer, Opt. Lett. {\bf 39}, 6624 (2014). 

\bibitem{VBPA15} M. L. Villinger, M. Bayat, L. N. Pye, A. Abouraddy, Opt. Lett. {\bf 40}, 5550 (2015).

\bibitem{BBFSN15}L. Baldacci, S. Zanotto, and A. Tredicucci, Rend. Fis. Acc. Lincei {\bf 26}, 219 (2015). 

\bibitem{SCCWBDPCMS16}B. C. P. Sturmberg, T. K. Chong, D-Y Choi, T. P. White, L. C. Botten, K. B. Dossou, 
C. G. Poulton, K. R. Catchpole, R. C. McPhedran, C. M. de Sterke, Optica {\bf 3}, 556 (2016).

\bibitem{STLLC14}Y. Sun, W. Tan, H.-q. Li, J. Li, H. Chen, Phys. Rev. Lett. {\bf 112}, 143903 (2014).

\bibitem{S50}J. Slater, Microwave Electronics (Van Nostrand, Princeton,1950).

\bibitem{LSMSP08} N. I. Landy, S. Sajuyigbe, J. J. Mock, D. R. Smith, and W. J. Padilla, Phys. Rev. Lett. {\bf 100}, 207402 (2008).

\bibitem{PL10} W. Padilla and X. Liu, SPIE Newsroom (2010), http://www.spie.org/newsroom/3137-perfect-electromagnetic-absorbersfrom-microwave-to-optical.

\bibitem{PPVZRKL13} V. T. Pham, J.W. Park, D. L. Vu, H. Y. Zheng, J. Y. Rhee, K.W. Kim, and Y. P. Lee, Adv. Nat. Sci. Nanosci. 
Nanotechnol. {\bf 4}, 015001 (2013).

\bibitem{DK38}W. Dallenbach and W. Kleinsteuber, Hochfrequenztechnik und Elektroakustik {\bf 51}, 152 (1938).

\bibitem{S52} W.W. Salisbury, U.S. Patent No. 2,599,944, 10 June (1952).

\bibitem{MMYYWS12}J. Mei, G. Ma, M. Yang, Z. Yang, W. Wen, and P. Sheng, Nat. Commun. {\bf 3}, 756 (2012).

\bibitem{SBHL14} J. Z. Song, P. Bai, Z. H. Hang, and Y. Lai, New J. Phys. {\bf 16}, 033026 (2014).

\bibitem{RTRMTP16} V. Romero-Garcia, G. Theocharis, O. Richoux, A. Merkel, V. Tournat, and V. Pagneux, Sci. Rep. {\bf 6}, 19519 (2016).

\bibitem{MYXYS14}G. Ma, M. Yang, S. Xiao, Z. Yang, P. Sheng, Nature Mater. {\bf 13}, 873 (2014).

\bibitem{CGMM13} F. Costa, S. Genovesi, A. Monorchio, and G. Manara, IEEE Trans. Antennas Propag. {\bf 61}, 1201 (2013).

\bibitem{PCZW13} Y. Pang, H. Cheng, Y. Zhou, and J. Wang, J. Appl. Phys. {\bf 113}, 114902 (2013).

\bibitem{SLLREK12} J. Schindler, Z. Lin, J. M. Lee, H. Ramezani, F. M. Ellis, and T. Kottos, J. Phys. A {\bf 45}, 444029 (2012).

\bibitem{LGJSY05} M. Law, L. Greene, J. Johnson, R. Saykally, P. Yang, Nat. Mater. {\bf 4}, 455 (2005).

\bibitem{T07}B. Tian, {\it et al.}, Nature (London) {\bf 449}, 885 (2007).

\bibitem{AP10}H. A. Atwater, A. Polman, Nat. Mater. {\bf 9}, 205 (2010)

\bibitem{LCG06}M. Laroche, R. Carminati, J.-J. Greffet, J. Appl. Phys. {\bf 100}, 063704 (2006)

\bibitem{LH08}A. Luque, S. Hegedus, {\it Handbook of Photovoltaic Science and Engineering} (Wiley, 2008).

\bibitem{FCDPRTTW00}M. Fink, D. Cassereau, A. Derode, C. Prada, P. Roux, M. Tanter, J.-L. Thomas, F. Wu, {\it Time-reversed acoustics}, 
Rep. Prog. Phys. {\bf 63}, 1933 (2000)

\bibitem{BPTB02}L. Borcea, G. Papanicolaou, C. Tsogka, J. Berryman, {\it Imaging and time reversal in random media}, Inverse Probl. {\bf 18}, 1247 (2002).

\bibitem{MRDNF02}G. Montaldo, P. Roux, A. Derode, C. Negreira, M. Fink, {\it Ultrasound shock wave generator with one-bit time reversal in a 
dispersive medium, application to lithotripsy}, Appl. Phys. Lett. {\bf 80}, 897 (2002).

\bibitem{CPCLD04}J. Dela Cruz, I. Pastirk, M. Comstock, V. Lozovoy, M. Dantus, {\it Use of coherent control methods through scattering biological tissue to achieve functional imaging}, Proc. Natl Acad. Sci. USA {\bf 101}, 17001 (2004).

\bibitem{HRY15}R. Horstmeyer, H. Ruan, C. Yang, {\it Guidestar-assisted wavefront-shaping methods for focusing light into biological tissue} , Nat. Photonics {\bf 9}, 563 (2015).

\bibitem{FM88}R. L. Fante, M. T. McCormack, IEEE Trans. Ant. Prop. {\bf 36}, 1443 (1998)

\bibitem{VJ96}K. Vinoy, R. Jha, {\it Radar absorbing materials - From theory to design and characterization} (Kluwer Academic Publishers, 1996).

\bibitem{M12} J. Mei, {\it et al.} Nature Communications {\bf 3}, 756 (2012).

\bibitem{ZFYZGSF16}H. Zhao, W. S. Fegadolli, J. Yu, Z. Zhang, L. Ge, A. Scherer, L. Feng, Phys. Rev. Lett. {\bf 117}, 193901 (2016)

\bibitem{LSFSK17}H. Li, S. Suwunnarat, R. Fleischmann, H. Schanz, T. Kottos, Phys. Rev. Lett. {\bf 118}, 044101 (2017).

\bibitem{FSK17}Y. V. Fyodorov, S. Suwunnarat, T. Kottos, J. Phys. A: Math. Theor. {\bf 50}, 30LT01 (2017).

\bibitem{LSK17}H. Li, S. Suwunnarat, T. Kottos, eprint arXiv:1712.00510 (2017)

\bibitem{Li2018} H. Li, T. Kottos, and B. Shapiro, Phys. Rev. Applied \textbf{9}, 044031 (2018). 

\bibitem{note0}We assume a fixed driving scheme (i.e. frequency and modulation) and solve for the corresponding
$(E_{\rm FPA},\gamma_{\rm FPA})$. Of course the calculations can be also performed assuming $\gamma$ fixed and
identifying the set $(E_{\rm FPA},\omega_{\rm FPA})$ for which PA occurs.  


\bibitem{PLF14}J. R. Piper, V. Liu, S. Fan, Appl. Phys. Lett. {\bf 104}, 251110 (2014)

\bibitem{LKS18}H. Li, T. Kottos, and B. Shapiro, Phys. Rev. A \textbf{97}, 023846 (2018).

\bibitem{note1}In the calculation we assumed a scenario for which the driving frequency $\omega$
is slightly larger than $2$ and thus $e^{-\imath k_{1}}\approx-\omega/2$.

\bibitem{H84}H.A. Haus, Waves and fields in optoelectronics (Prentice-Hall, Englewood Cliffs, NJ, 1984)

\bibitem{CLEK17}M. Chitsazi, H. Li, F. M. Ellis, T. Kottos, Physical Review Letters {\bf 119}, 093901 (2017)
\end{thebibliography}
\end{document}